\documentclass[12pt]{iopart}

\usepackage{graphicx}
\begin{document}

\title[Experimental study of charging of dust grains in presence of energetic electrons]{Experimental study of charging of dust grains in presence of energetic electrons}

\author{R. Paul$^{1}$, G. Sharma$^{1}$, K. Deka$^{1}$,S. Adhikari$^{2}$,R. Moulick$^{3}$, S. S. Kausik$^{1,*}$,B. K. Saikia$^{1}$}

\address{$^{1}$Centre of Plasma Physics, Institute for Plasma Research, Sonapur-782402, Kamrup (M), Assam, India.}%Lines break automatically or can be forced with \\
\address{$^{2}$Department of Physics, University of Oslo, PO Box 1048 Blindern, NO-0316 Oslo, Norway.}
\address {$^{3}$Department of Physics, Rangapara College, Rangapara, Sonitpur, Assam-784505, India.}

\address{$^*$E-mail: kausikss@rediffmail.com}
\vspace{10pt}

\begin{abstract}
The role of hot electrons in charging of dust grains is investigated in a two temperature hydrogen plasma.  A variety of dust particles are introduced into the system and secondary electron emission (SEE) from each of the dust grains has been reported.  A cylindrical Langmuir probe is used for determining the plasma parameters and a Faraday cup is connected to an electrometer in order to measure the dust current. The electrometer readings confirm the electron emission from the dust and SEE is observed from the tungsten dust in a low-pressure experimental plasma device for the first time.
\end{abstract}
%
% Uncomment for keywords
%\vspace{2pc}
%\noindent{\it Keywords}: XXXXXX, YYYYYYYY, ZZZZZZZZZ
%
% Uncomment for Submitted to journal title message
%\submitto{\JPA}
%
% Uncomment if a separate title page is required
%\maketitle
% 
% For two-column output uncomment the next line and choose [10pt] rather than [12pt] in the \documentclass declaration
%\ioptwocol
%

\section{\label{intro}Introduction}

Charging of dust grains has altogether been a fascinating field of  research for several decades. The ability of dusty plasmas to explain the various astrophysical phenomena as well as its relevance in the industrial applications establishes it as a potential field of research \cite{shukla}. Among the numerous known mechanisms of dust charging, electron impact ionization is the most probable agent in low pressure plasmas. In  such plasmas, temperature of the incident electrons plays a pivotal role in charging processes. Increase in the temperature increases the kinetic energy of the electrons, thereby enhancing the current contribution on the dust grains \cite{kausik}. Usually, a high energy tail in  the electron energy distribution is observed in  low pressure laboratory plasmas \cite{nakamura_komatsuda_1998}. In the presence of  such an energetic electron group, dust charging scenario may change significantly. The advent of fusion research has led to the quest for a material that can sustain in such an extreme environment. Research shows that tungsten(W) may be the most promising material in this regard \cite{PITTS2013S48, HIRAI2016616}. Exposing W to high heat flux region of tokamak might result in generation of nano to micron sized dust particles in the system which is often regarded as contaminants \cite{SARMAH2018120}.  Therefore, understanding the dust charging mechanism becomes essential for removing them from the fusion devices \cite{shukla}. Present study investigates the response of micron-sized W dust in presence of energetic electrons in a low pressure plasma environment. In  fusion devices, it is usually seen that there is a fuel loss due to the retention of the fuel (H$^+$) by the dust \cite{SARMAH2018120}. This is a prime concern for the fusion devices, which may be avoided if the dust present in the device \cite{SARMAH2018120}  is positively charged \cite{Winter_2004}. The present work establishes an alternative and a simpler method of reducing negative charges on the dust surface which may help in producing positively charged dust. The plasma environment for the present work is comparable to the outer strike point and low-field side of the divertor region of the COMPASS tokamak \cite{Hasan_2018}, hence this study might be useful in overcoming the dust related complexities often encountered in such fusion devices. To verify the effectiveness of the present method, W dust has been replaced by alumina (Al$_2$O$_3$), as earlier studies confirms SEE from Al$_2$O$_3$ \cite{kakati}. Moreover, cesium coated W dust is also chosen in this study due to its relatively low work function than W and Al$_2$O$_3$ dusts. It also plays a noteworthy role in producing negative ions which are then used for NBI systems in the fusion devices \cite{bharat,Ikeda_2006}. \newline 

The paper has been divided into the following sections, Sec. \ref{apparatus} discusses the apparatus and the experimental system. Sec. \ref{results} analyze the experimental findings and a brief conclusion has been presented in Sec. \ref{conclusion}.

\section{\label{apparatus} Apparatus and Experimental Setup} 

\begin{figure}
    \centering
    \includegraphics[width=0.5\textwidth]{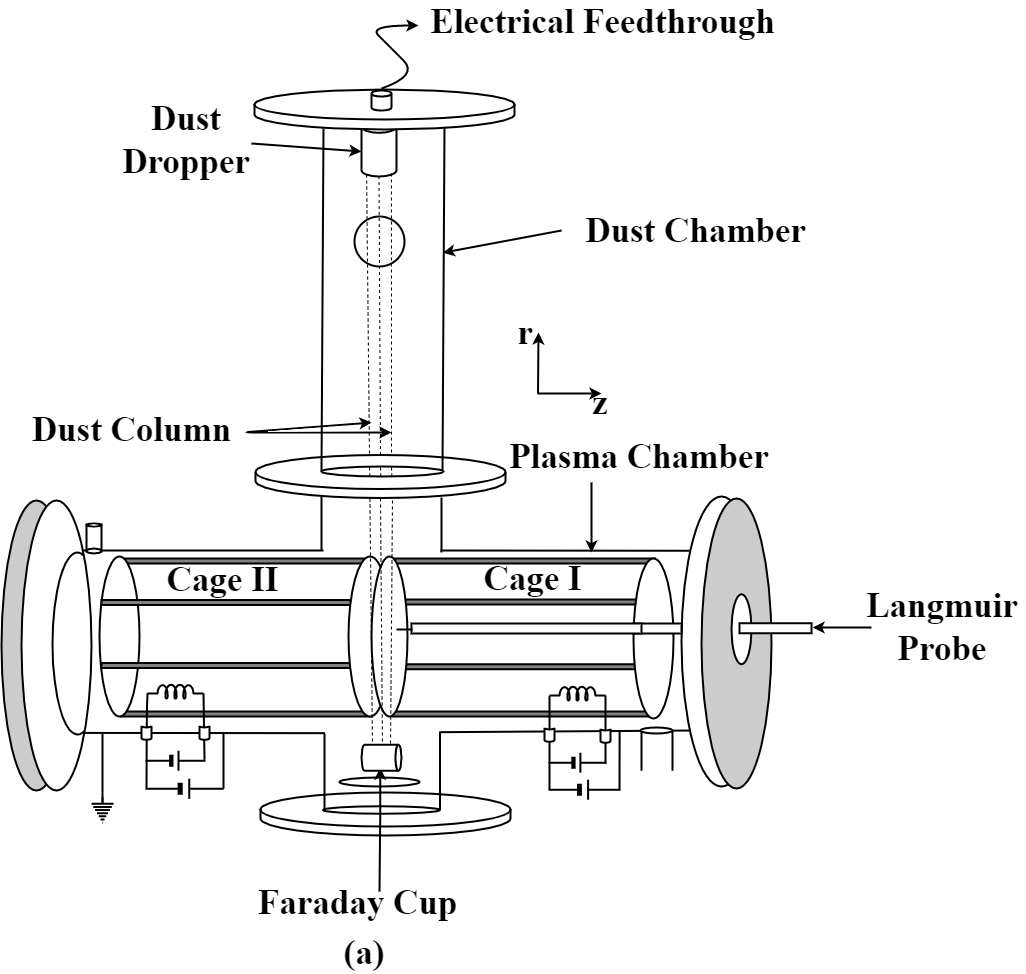}
    \caption{Schematic diagram of the experimental set up.}
    \label{fig:circuit}
\end{figure}

A two-temperature electron plasma is produced by a hot cathode discharge method.
The experimental set-up consists of two stainless steel chambers placed one above the other, as shown in figure \ref{fig:circuit}. The lower chamber placed horizontally is the plasma chamber, and the upper chamber placed vertically holds the dust dropper. The plasma chamber is  $100$ cm in length and $30$ cm in diameter. The dust dropping unit, on the other hand, has a height of $72$ cm and $15$ cm in diameter. The chamber is evacuated by a pumping arrangement consisting of a diffusion pump ($1000$ lit/s) backed by a rotary pump ($600$ lit/min). \newline

\begin{figure}
    \centering
    \includegraphics[width=0.5\textwidth]{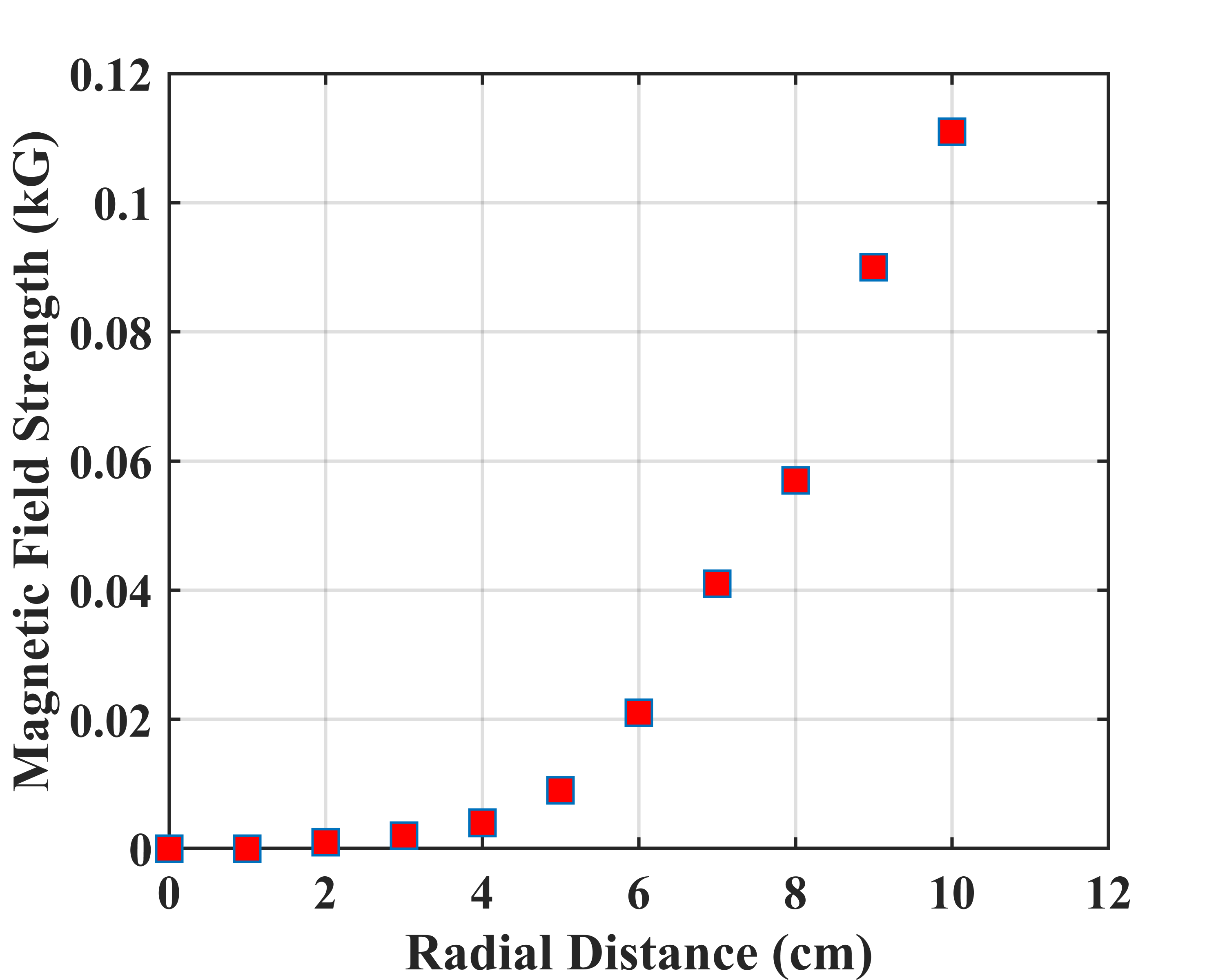}
    \caption{Gauss meter readings of the radial magnetic profile (from centre to the periphery) at the junction of Cage I and Cage II.}
    \label{fig:gauss}
\end{figure}

Two cylindrical magnetic cages of different field strengths are used for confining the plasma. Cage I  has rectangular shaped samarium cobalt magnets (surface magnetic field of $3.5$ kG), filled inside twelve stainless steel channels  and requires a water cooling system. The total length of the magnetic cage is 40 cm and the outer diameter is 28 cm.  Cage II on the other hand, has cubical shaped strontium ferrite magnets (surface magnetic field strength of $1.2$ kG) inside ten stainless steel channels. The total length of Cage II equals that of Cage I but the outer diameter is 29 cm. The Gauss meter reading, as shown in figure \ref{fig:gauss}  confirms that the junction of the two cages is magnetic field free. On moving from the centre to the periphery of the junction, a nominal increase can be observed in the magnetic field strength.  Thoriated tungsten filaments are used as a source of primary electrons in the system.  The electrons emitted from the filaments are accelerated by applying a  $80$ V discharge voltage. Plasma is produced by striking a discharge between the filaments and the grounded magnetic cage. A Hiden Analytical Limited's ESPION advanced Langmuir probe system is used for determining the plasma parameters at the junction.\newline
\begin{figure}
    \centering
    \includegraphics[width=0.5\textwidth]{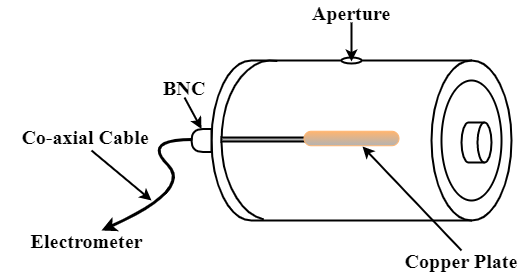}
    \caption{Schematic diagram of Faraday cup.}
    \label{fig:faraday_cup}
\end{figure}

Dust particles are introduced in the plasma using a dust dropping unit placed above the horizontal chamber. The chosen dusts are  almost spherical in shape and have  similar radii ($1.5$ $\mu m$ ). The dust dropping unit comprises of a dust dropper connected externally to an oscillator circuit used for vibrating the dropper. The dust dropper is made up of a container with a stainless steel mesh as the base and same mesh is used for dropping the dust grains inside the system. The vibrating frequency can be varied by controlling the input voltage applied from an external power supply.  Al$_2$O$_3$  dust is lighter than W dust, and a suitable input voltage and frequency is applied to the dust dropper to maintain a uniform dust density inside the chamber. For all the samples the dust density is of the order of  $10^5/cm^3$ measured using the laser scattering technique \cite{nakamura}.  The dust particles interact with the diffused plasma and eventually gets charged. Charging time of the dust \cite{cui,goree} inside the plasma in case of a single cage operations is approximately equal to a few microseconds depending upon the plasma density and electron temperature \cite{kakati_2014}. On the other hand, the time taken by the dust particle to cross the plasma volume under gravity is approximately equal to few milliseconds \cite{kakati_2014}. Thus, the dust grains have enough time to attain the equilibrium charge state within the plasma volume.  For measuring the dust current, a sensitive electrometer (Keithley Instruments, 6514) is used. It is attached to a Faraday cup with a low noise tri-axial cable as shown in figure \ref{fig:faraday_cup}. The Faraday cup assembly sits exactly below the junction of the two cages and is separated by a distance of approximately 5 cm from them. Two permanent magnets (surface field strength 1.2 kG each) are axially placed across the Faraday cup which creates a strong transverse magnetic field.  \newline
\begin{figure}[ht] 
  \begin{minipage}[b]{0.45\linewidth}
    \centering
    \includegraphics[width=1.0\textwidth]{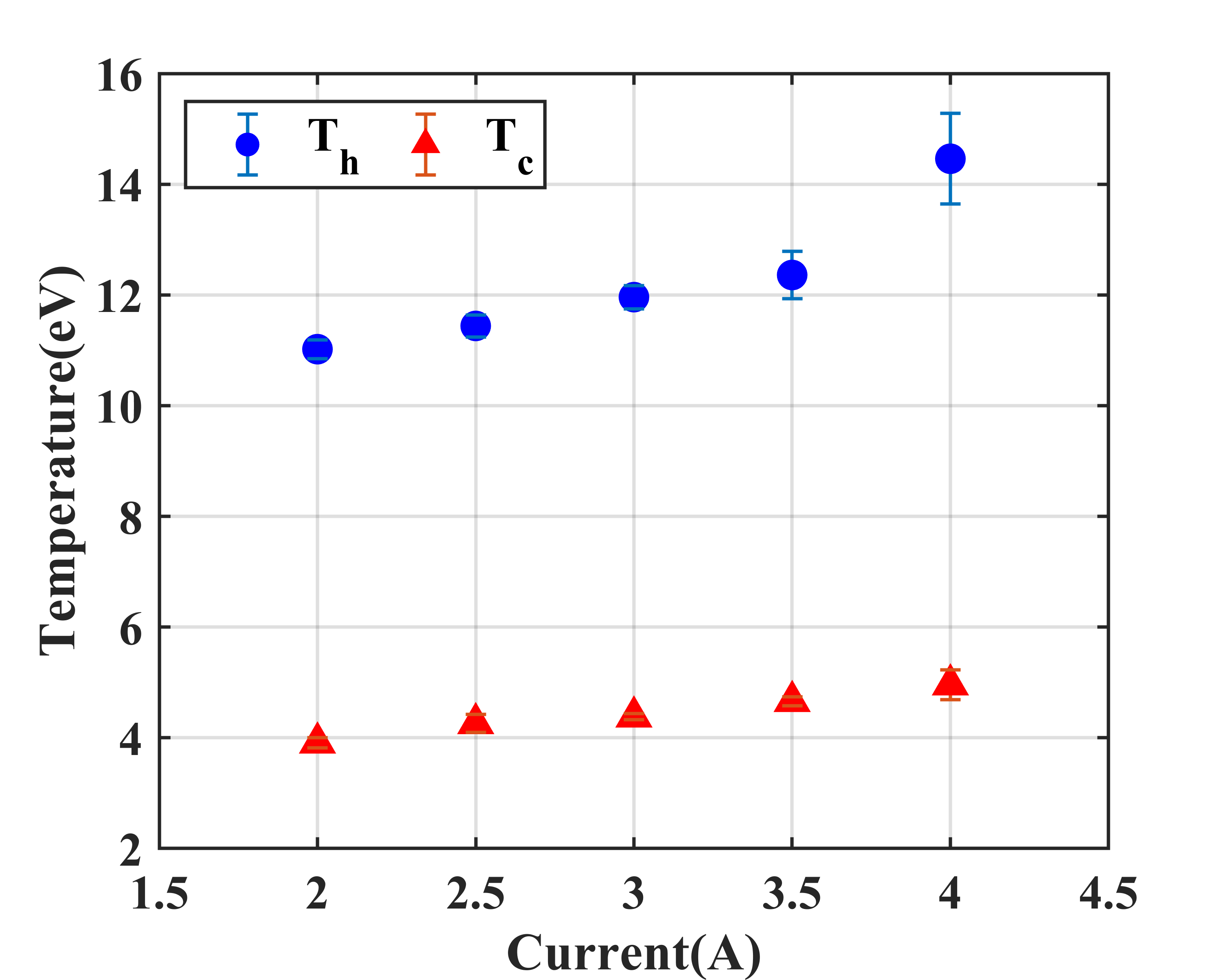}
    \caption{Temperature of the electron groups for different discharge currents in Cage I and a constant 0.2 A discharge current in Cage II.}
     \label{fig:temp_1}
    \vspace{4ex}
  \end{minipage}%%
  \begin{minipage}[b]{0.45\linewidth}
    \centering
    \includegraphics[width=1.0\textwidth]{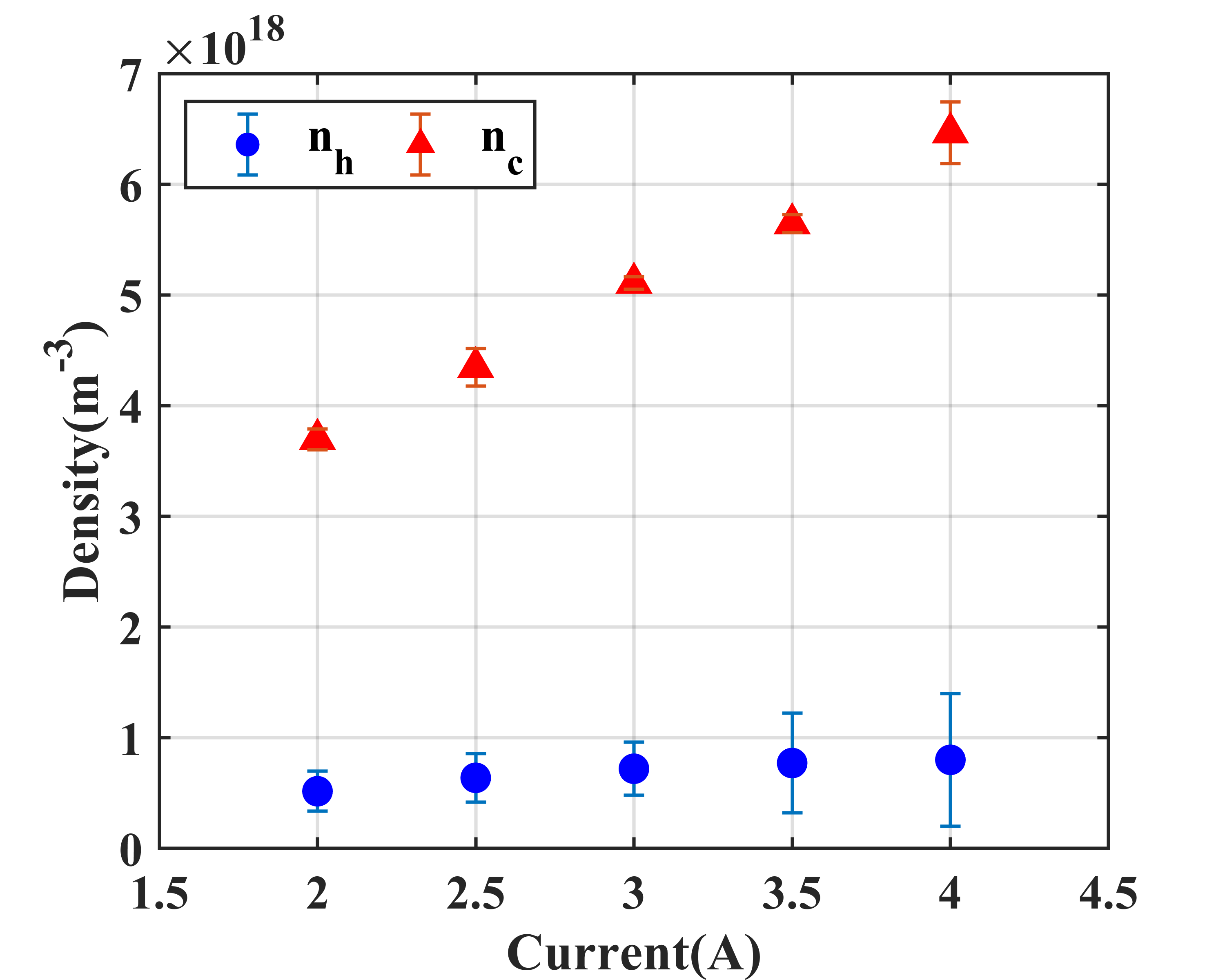}
    \caption{Densities of the electron groups for different discharge currents in Cage I and a constant 0.2 A discharge current in Cage II.}
    \label{fig:temp}
    \vspace{4ex}
  \end{minipage} 
  \end{figure}

\section{\label{results} Results}

\begin{figure}
    \centering
    \includegraphics[width=0.5\textwidth]{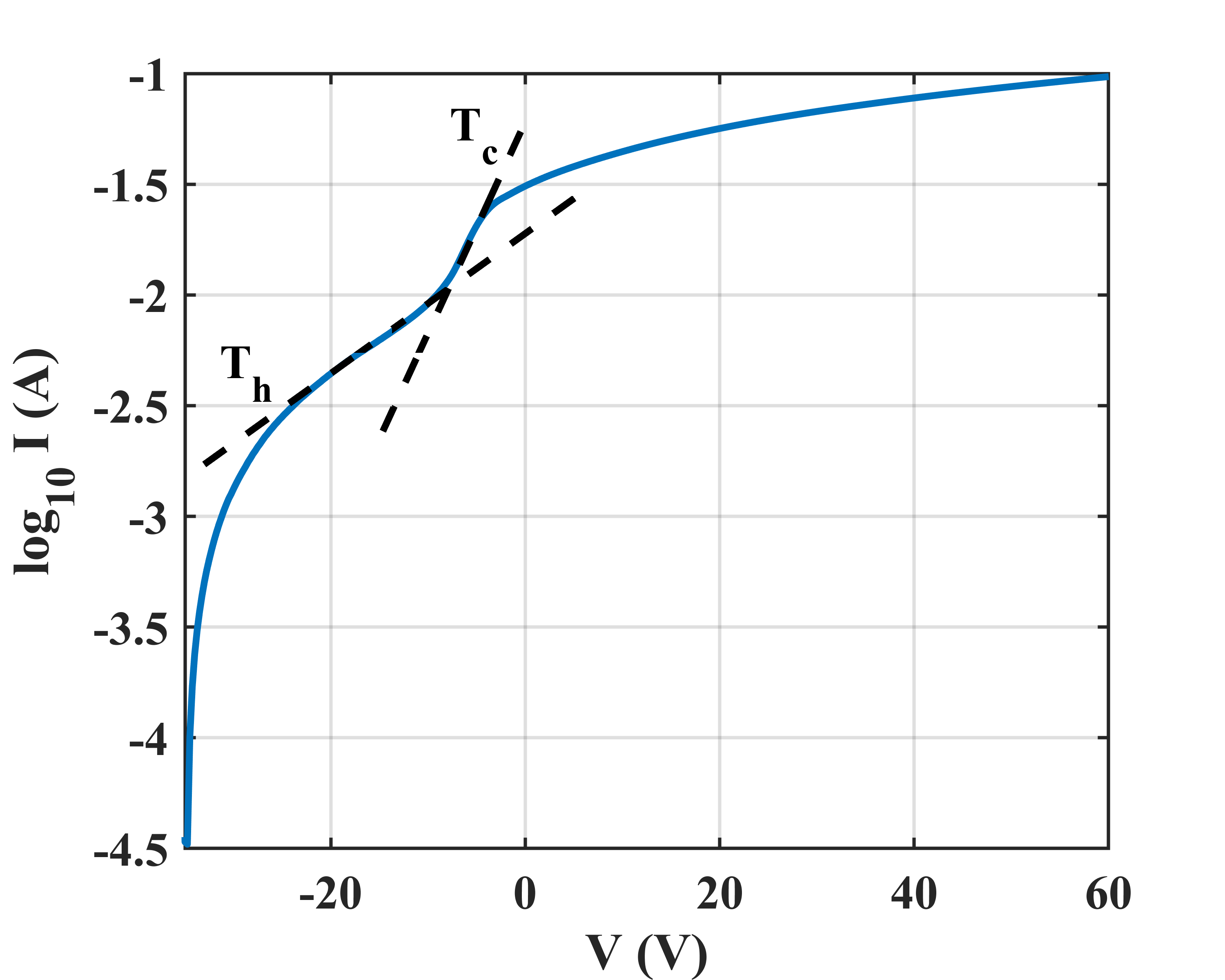}
    \caption{Semi-logarithmic plot of the \textit{I-V} curve for a discharge current 4 A in Cage I and 0.2 A in Cage II.}
    \label{fig:i-v}
    \end{figure}

\begin{figure}
    \centering
    \includegraphics[width=0.5\textwidth]{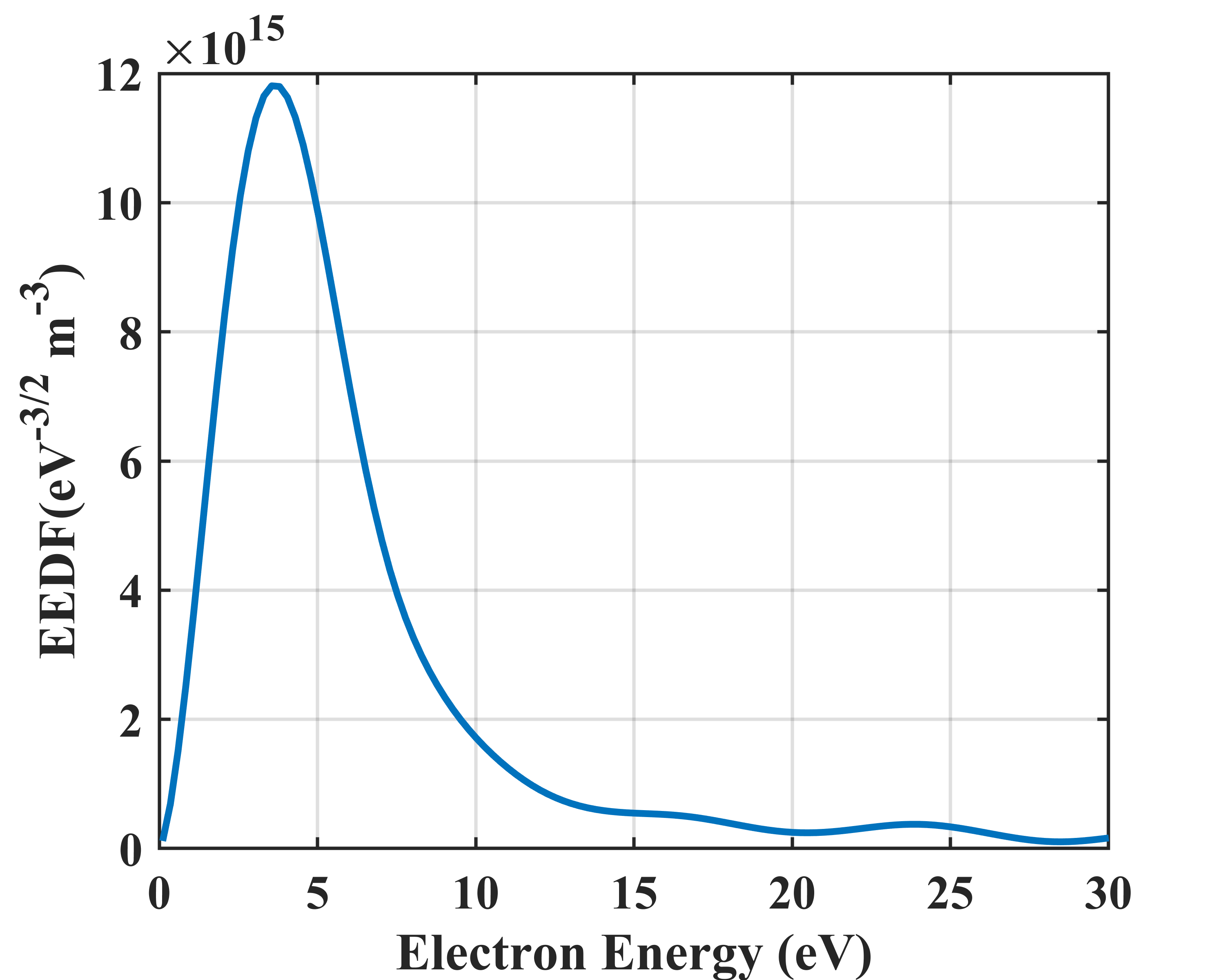}
    \caption{ EEDF plot for a discharge current 4 A in Cage I and 0.2 A in Cage II.}
    \label{fig:EEDF}
       \end{figure}

\begin{figure}
    \centering
    \includegraphics[width=0.5\textwidth]{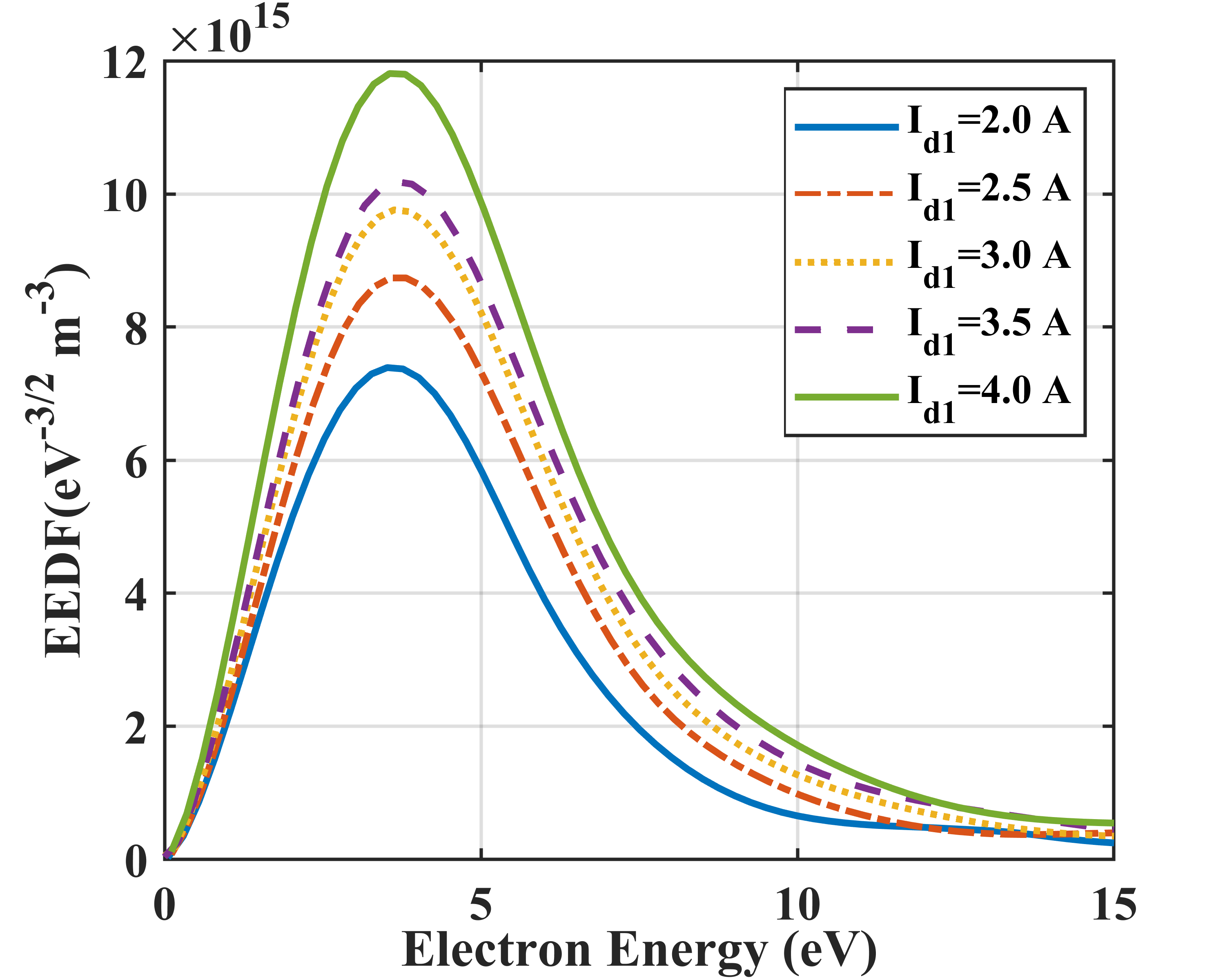}
    \caption{  EEDF for different discharge currents in Cage I (I$_{d1}$) and a constant 0.2 A discharge current in Cage II.}
    \label{fig:eedf}
\end{figure}

Hydrogen plasma is produced through a hot cathode discharge method inside the two individual cages maintained at a constant working pressure of $2\times10^{-4}$ mbar, thereafter it diffuses at the edge of the cages. The diffused plasma so formed at the junction contains two electron groups of distinct temperatures and densities as seen in figures \ref{fig:temp_1} and \ref{fig:temp}. Traces of two electron groups can also be observed in the semi-logarithmic plot of the \textit{I-V} curve as shown in figure \ref{fig:i-v}.  Additionally, EEDF is calculated and is shown in figure \ref{fig:EEDF}.  Usually, in single cage operation, it is seen that a high density plasma is obtained in Cage I due to its higher confining ability, while, Cage II has a low density plasma due to its lower surface field strength. Kakati \textit{et al.}  \cite{kakati_2014} have found a high energy tail for a discharge at a lower pressure ($10^{-5}$ mbar) which gradually fades away with the increase in the gas pressure. However, the present arrangement ensures that for a pressure  of $10^{-4}$ mbar the high energetic tail of the EEDF is retained. By maintaining a higher discharge current in Cage I and that of a lower one in Cage II provides an optimum condition for attaining two electron population in the system. Hence, the discharge current in Cage II is kept constant at $0.2$ A and that in Cage I is varied from $2$ to $4$ A. Figure \ref{fig:eedf} shows the EEDFs for various discharge conditions. It can be seen that as the discharge current in Cage I is increased, the high energy tail shifts towards right indicating an increase in the hot electron population. Also, with the increase in discharge current, the number of thermionic electrons emitted from the filaments increases thereby enhancing the degree of ionization. 
Due to this increase in the primary population, the effective electron temperature increases which is also accompanied with an increase in the electron density as observed from figure \ref{fig:eedf}. The two temperature electron plasma production mechanism has been explained and the details of which can be found here  \cite{gunjan}. 
\newline 

\begin{figure}
    \centering
    \includegraphics[width=1.0\textwidth]{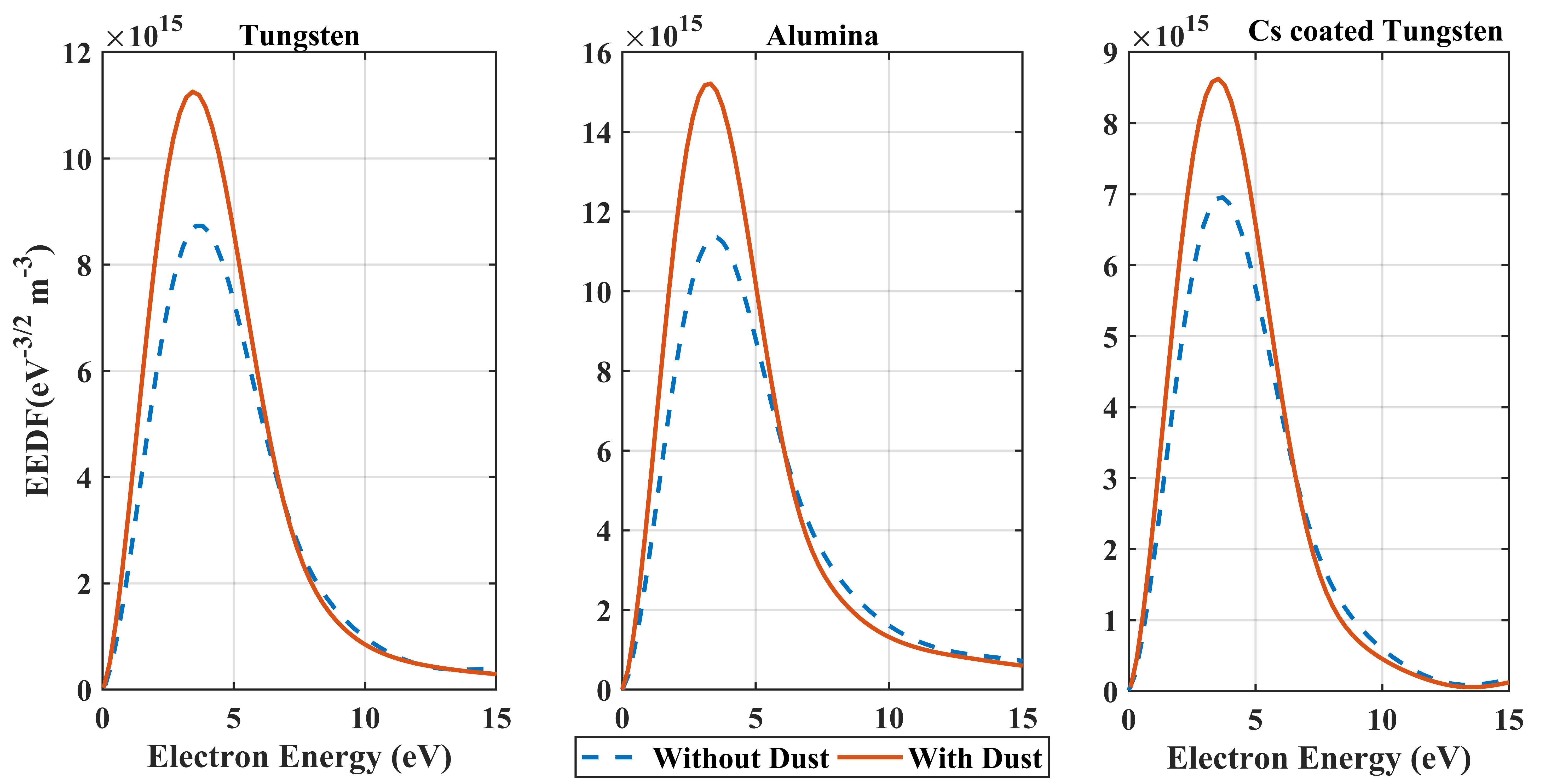}
    \caption{ EEDF for different dust samples for a constant discharge current 2.5 A in Cage I and 0.2 A in Cage II.}
    \label{fig:Dust_EEDF}
\end{figure}

\begin{figure}
    \centering
    \includegraphics[width=0.5\textwidth]{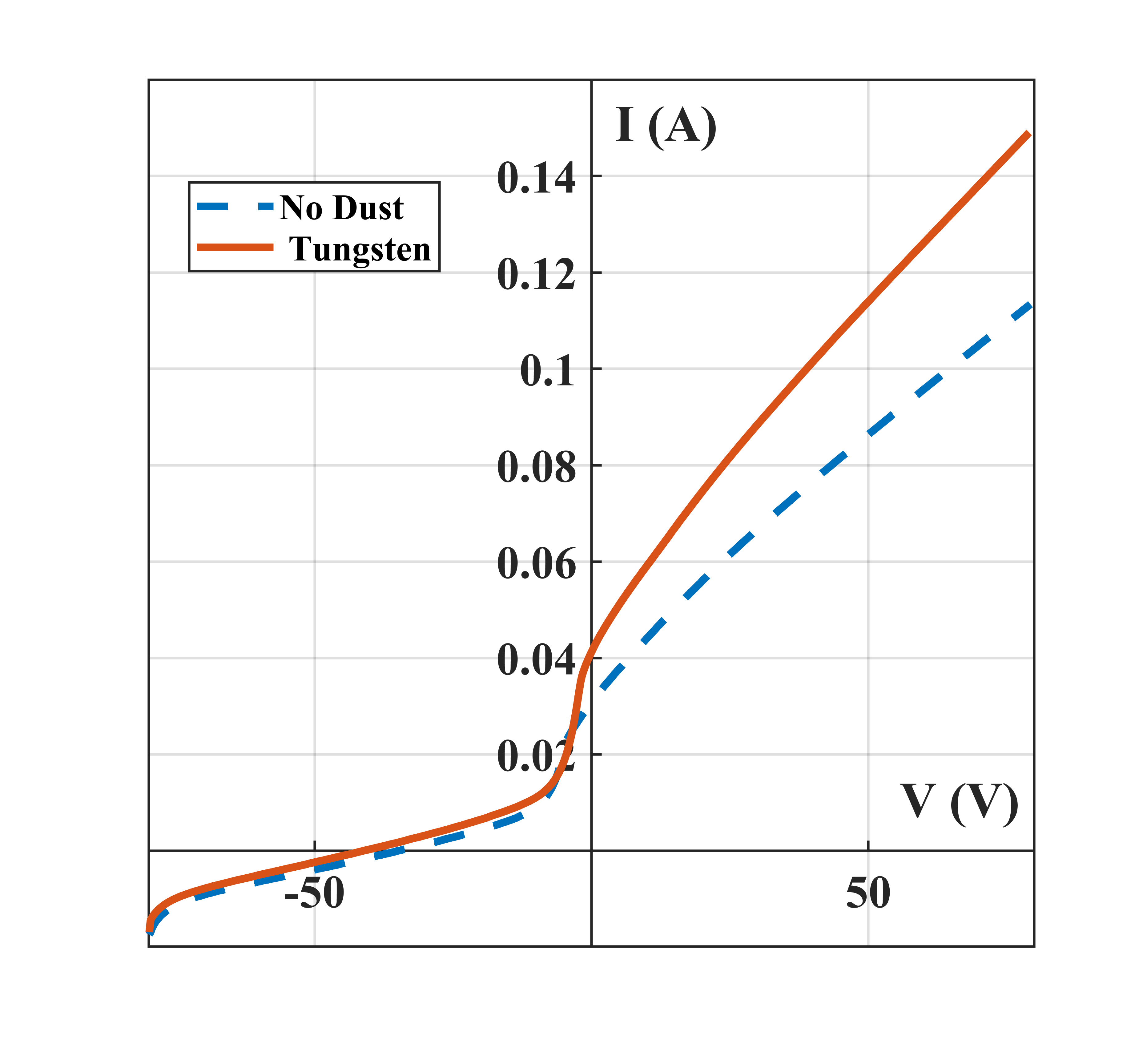}
    \caption{ \textit{I-V} curve in presence and absence of W dust for a discharge current 4 A in Cage I and 0.2 A in Cage II.}
    \label{fig:Dust_I-V}
\end{figure}

With the introduction of dust in the plasma, it is observed from the previous study  \cite{kakati_2014} that the higher energy tail of the EEDF increases and the bulk population peak decreases. The mid energy electrons get attached to the dust, results in a decrease in the bulk electron population. Due to this reason, the overall electron saturation current decreases in presence of dust and was first reported by Barkan \textit{et al.} \cite{Barkan}. But, a completely different scenario has been observed in the present study. EEDFs in presence of different dust particles at the same plasma conditions has been calculated and is shown in figure \ref{fig:Dust_EEDF}. In presence of dust, there is a significant rise in the bulk electron population along with a left hand shift in the EEDF and a depletion in the high energy tail is observed. In addition, the overall electron saturation current increases in presence of dust and can be seen from the \textit{I-V} curve in figure \ref{fig:Dust_I-V}.  This phenomenon is observed for all the chosen dust samples. Among them Al$_2$O$_3$ has the highest SEE yield (in the range of 2 to 3) for the primary electrons in the energy range of $60-100$ eV \cite{kakati} as compared to the others. Due to this reason, SEE from alumina dust has been observed earlier in a single cage operation \cite{kakati}. However, tungsten (W) dust has a comparatively lower SEE yield (in the range of 0.5) for the above mentioned energy range. Hence, SEE was not observed for W dust in the previous study \cite{kakati}. Although, the primary electrons, in the present arrangement, fall within the same energy range, it is observed that W dust also exhibits SEE. \newline

There are many well known possible mechanisms through which a dust grain can acquire charge in a plasma environment \cite{shukla}.
 For the present case, it appears that the most suitable means by which dust can be charged is its interaction with the energetic particles. Among the plasma species present in the system,  electrons are only responsible for charging the dust particles. Usually, in low pressure laboratory plasmas, the lower energetic electrons are in bulk as compared to the hot electrons. This leads to the attachment of these cold electrons on the dust surface. Consequently, dust grains becomes negatively charged. SEE can occur in situations, where the hot electron population is significant and have energy above 10 eV, such electrons can tunnel through the dust surface and initiate SEE   \cite{chow1,chow2}. When they strike the particles, a part of their energy is lost to the dust. The lost energy is used in exciting the electrons as a result of which an emission of electrons take place from the dust grain \cite{mamun}. The emitted electrons have energy comparable to that of the cold electrons. Due to this reason, the low energy electron population increases significantly and is reflected in the EEDF plots as shown in figure \ref{fig:Dust_EEDF}. On the other hand, a depletion in the hot population is also observed in the EEDF plot in figure \ref{fig:Dust_EEDF} indicating an absorption of the energetic electrons by the dust grains. Goertz \cite{goertz} predicted that the emitted electrons will have a Maxwellian distribution and will lie in the energy range between $1-5$ eV. This is reflected in the EEDFs curves as shown in figure \ref{fig:Dust_EEDF}. Therefore, in the present scenario, the deviation from the familiar theoretical result for the observed dust current with the increasing electron temperature might be credited to secondary electrons emitted from the dust surface \cite{susmita}. Cesium coated W dust grains also show similar SEE behavior as seen in figure\ref{fig:Dust_EEDF}. \newline

\begin{figure}
    \centering
    \includegraphics[width=0.5\textwidth]{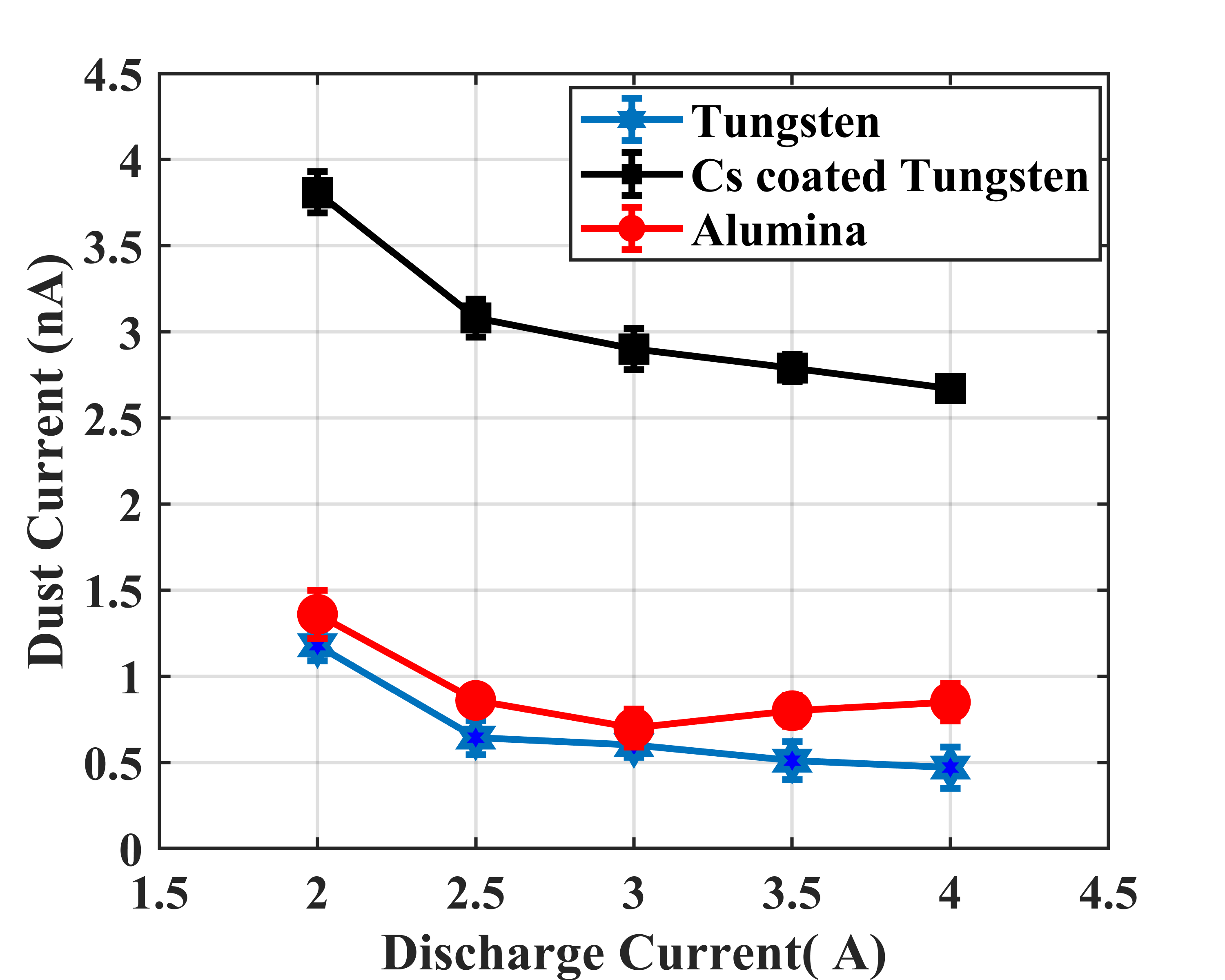}
    \caption{ Dust current readings for W, Cs coated W and Al$_2$O$_3$ dusts for different discharge currents in Cage I and a constant 0.2 A discharge current in Cage II.}
    \label{fig:Dust_Current}
\end{figure}

The variation of dust current for W, Cs coated W and  Al$_2$O$_3$ dusts with varied discharge current is shown in figure \ref{fig:Dust_Current}. Usually, dust becomes negative by absorbing electrons from the ambient plasma for which the electrometer readings indicate a negative dust current confirming electron absorption by the dust grains \cite{kakati}. However, an opposite behavior is observed in this particular study as seen from figure  \ref{fig:Dust_Current}. Here, the dust current measured for all the dust samples  consequently refers to an emission of electrons from the dust grains. This decrease in the dust current also confirms occurrence of SEE from the dust surfaces. This indicates that the negative charge accumulated on the dust surface decreases with the increase in the discharge current.  Many theoretical findings \cite{susmita,gupta,bhakta,miki} suggest that positively charged dust is a stable state that arises due to the emission of electrons from the dust surface. D'Angelo \cite{ANGELO} measured positively charged dust in presence of negative ions experimentally. A similar method has been adopted by Kim and Merlino \cite{kim} for obtaining positively charged dust. But, the present method is quite different from the above stated techniques. A comparison of dust currents between the dust samples as shown in figure \ref{fig:Dust_Current} reveals that Cs coated W dust emits a larger fraction of electrons than that of pure W and Al$_2$O$_3$ dusts. The reason lies in the fact that Cs layer on W lowers the work function \cite{bharat} (from 4.5eV to 1.8 eV) facilitating in a  higher emission rate. Al$_2$O$_3$ dust on the other hand, has a higher SEE yield as compared to W dust, so the electrons are emitted easily from the dust surface which is reflected by a higher dust current in comparison to W. The rate of decrease in the measured dust current with respect to the increase in the discharge current is maximum for Cs coated W, followed by Al$_2$O$_3$ and W. Since, decrease in the dust current consequently refers to electron emission from the dust surface, hence confirms the EEDF findings and validates our assumption about SEE. \newline    

\begin{figure}
    \centering
    \includegraphics[width=1.0\textwidth]{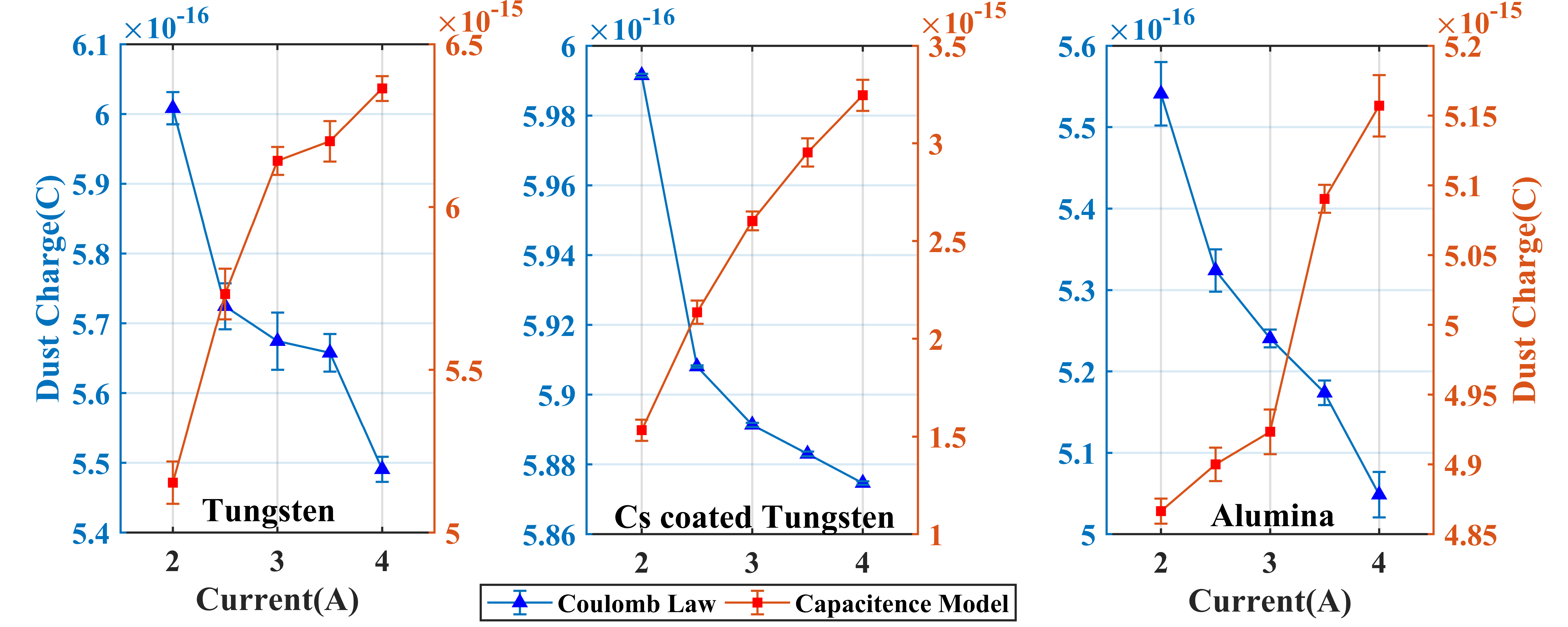}
    \caption{ Dust charge calucation for various dust samples for different discharge currents in Cage I (I$_{d1}$) and a constant 0.2 A discharge current in Cage II.}
    \label{fig:Dust_Charge}
\end{figure}

Since the chosen dusts are spherical in shape, the charge accumulated on the dust is often calculated by the capacitance model. When the net current on the dust grain vanishes, the dust surface acquires a floating potential ($\phi_p$). At this equilibrium state, the floating potential is related to the charge on the dust grain ($q$) as
\begin{equation}
  q = 4 \pi \epsilon_0  r_d \phi_p,  
\end{equation}
where $r_d$ is the radius of the chosen dust grain. The dust surface potential, $\phi_p$ is calculated as
\begin{equation}
  \phi_p= V_f - V_p,   
\end{equation}
where $V_f$ and $V_p$ are the respective floating and plasma potentials.
This theoretical model is useful in calculating the amount of charge gathered on the dust. But, in case of secondary emissions, as this model does not incorporate electron emissions, hence it becomes inadequate for such situations \cite{b.k}. Therefore, a new model \cite{austin} has been adopted in this study for calculating the dust charge. According to this model, the peaks in the EEDFs in presence of dust corresponds to the floating potential($\Phi$). According to the theoretical solution for EEDF, as the emission takes place from the dust, electrons are injected into the plasma with an energy comparable to the dust floating potential. As the observed peaks are due to the emitted electrons from the dust samples, hence the corresponding potential depicts the actual floating potential of the dust. This potential is used in calculating the dust charge ($Q$) from the Coulomb's law given by
\begin{equation}
  \Phi= Q/(4\pi\epsilon_0 r_d),  
\end{equation}

where $r_d$ is the radius of the chosen dust grain.
A comparison is drawn between these two models and is shown in figure \ref{fig:Dust_Charge} for all the dust samples. The capacitance model shows an increasing trend for all the dusts with the increase in the discharge current. But this result is contrary to the electrometer findings as shown in figure \ref{fig:Dust_Current}  where dust current decreases with increase in discharge current. On the other hand, the charge calculated using the Coulomb law model shows a decreasing trend which resembles the dust current  readings. This decrease in the dust charge consequently stands for less negatively charged dusts measured by the electrometer. The decrease in the negative charge symbolizes emission of electrons from the dust thereby indicating SEE from the dust grains.\newline 
Two possible scenario arises due to the emission of electrons from the dust grains \textit{viz.}, when the incoming electron flux is greater than the outgoing flux and when the situation reverses. For the first case SEE yield ($\delta$) is less than unity and for the latter $\delta$ is greater than unity. An estimation of $\delta$ can be made from the following equation \cite{martin}
\begin{equation}
   (1-\delta)exp(-V_d)=\sqrt{\left(\frac{\theta}{\mu}\right)} \left(1+\frac{V_d}{\theta}\right), 
\end{equation}

where $V_d=\frac{-e\phi_d}{k_BT_e}$, $\theta=\frac{T_i}{T_e}$ and $\mu=\frac{m_i}{m_e}$. The dust potential $\phi_d$ is calculated in the similar manner as that of the already mentioned $\phi_p$. For the present case, the estimated $\delta$ value for W dust from the above equation is 0.9 which is more than the maximum $\delta$ value (0.5) for the electrons in the energy range 60-100 eV. This indicates that significant amount of electrons have been emitted from the dust surface. Since $\delta < 1$, the dust charge may not be completely positive in the present case, but nevertheless the negative charge on the dust is found to be decreasing. It has been observed that the parameter $\theta$ plays an important role in determining the value of $\delta$. Since $T_e >> T_i$ for the present system, hence a small change in $T_e$ has a significant effect on $\delta$. Moreover, the arrangement of the magnetic cages in this case, is seen to have an added advantage in enhancing the hot electron population. Since, the magnetic field lines are parallel to the axis of the chamber, this guides the electrons towards the centre of the device thereby, increasing the total electron flux at the centre significantly. As the dust falls at the junction of the two cages, the probability of the dust grains interacting with the energetic electrons becomes maximum. These impinging electrons have sufficient energy to initiate SEE from the dust grains. It can also be seen as a technique for reducing hot electron component in the system. Further, this method may be helpful in enhancing surface assisted volume production of negative ions in a plasma \cite{bharat}. \newline

\section{\label{conclusion}Conclusion}
The present work investigates the charging mechanism of the dust grains in presence of an energetic electron group. These hot electrons are capable of initiating SEE from dust surfaces as observed in the present experiment which is reflected in the \textit{I-V} curve. It is believed that the present orientation of the confining magnetic field has a profound effect in the dust charging mechanism. To our knowledge, SEE from W dust surface in the energy range of $60-100$ eV is  observed for the first time in a low pressure laboratory plasma.

\section*{Acknowledgements}
The authors are grateful to Mr. G. D. Sarma for his technical assistance during the experiment. The authors would also like to thank Dr. Nipan Das not only for his technical assistance but also for his valuable suggestions at different stages of the experiment.\\

\end{document}